# Shape Evolution of CdSe Nanoparticles controlled by Halogen Compounds

Michaela Meyns,[a]* Fabiola Iacono,[b, c] Cristina Palencia,[c,]† Jan Geweke,[a] Mauricio D. Coderch,[d] Ursula E.A. Fittschen,[d] José M. Gallego,[c,e] Roberto Otero,[b, c] Beatriz H. Juárez,[c, f]* Christian Klinke[a]*

a Institute of Physical Chemistry, University of Hamburg, Grindelallee 117, 20146 Hamburg, Germany

b Dpto. de Física de la Materia Condensada and Instituto Nicolás Cabrera, Facultad de Ciencias, Universidad Autónoma de Madrid, Avd. Fco. Tomás y Valiente 7, 28049 Madrid, Spain.

c IMDEA Nanoscience, Faraday 9, 28049 Cantoblanco, Campus de Cantoblanco, Madrid, Spain

d Institute of Inorganic and Applied Chemistry, University of Hamburg, Grindelallee 117, 20146 Hamburg, Germany

e Instituto de Ciencia de Materiales de Madrid, ICMM, CSIC, Sor Ángela de la Cruz s/n, 28049 Madrid

f Departamento de Química-Física Aplicada, Universidad Autónoma de Madrid, Avd. Fco. Tomás y Valiente 7, Cantoblanco 28049, Madrid, Spain

**ABSTRACT:** Halogen compounds are capable of playing an important role in the manipulation of nanoparticle shapes and properties. In a new approach, we examined the shape evolution of CdSe nanorods to hexagonal pyramids in a hot-injection synthesis under the influence of halogenated additives in the form of organic chlorine, bromine and iodine compounds. Supported by DFT calculations, this shape evolution is explained as a result of X-type ligand coordination to sloped and flat Cd-rich facets and an equilibrium shape strongly influenced by halides. Synchrotron XPS measurements and TXRF results show that the shape evolution is accompanied by a modification in the chemical composition of the ligand sphere. Our experimental results suggest that the molecular structure of the halogenated compound is related to the degree of the effect on both rod growth and further shape evolution. This presents a new degree of freedom in nanoparticle shape control and highlights the role of additives in nanoparticle synthesis and their possible *in situ* formation of ligands.

## 1. Introduction

The application of the "hot-injection method" contributed strongly to the synthesis of monodisperse spherical semiconductor nanoparticles (NPs).[1] From spherical structures scientific interest moved towards more complex geometries (rods, tetrapods, octapods)[2,3] and later their epitaxial combinations with further materials,[4, 5] higher dimension nanomaterials,[6,7] and hierarchical superstructures.[8,9] The basis of a successful formation of such sophisticated nanostructures is a high degree of control over the nucleation and growth mechanisms and synthetic parameters. Physicochemical properties and reactivity of NPs are determined to a high degree by their shape and thus their surface facets.[10] A general approach to tune NP shapes is to adjust the types of ligands and their relative concentration so that certain crystallographic facets are favored during growth.[11] Further methodologies of shape control include self-assembly processes of NPs in solution, resulting in 2D materials, for instance,[7, 12] and two-phase shape control where already formed NPs undergo a morphological transformation due to the addition of compounds which etch and/or protect specific crystallographic facets. This method is often employed in the synthesis of metal NPs.[13,14] For semiconductor NPs, oxidation-driven post-synthetic chemical etching or chemical and photochemical processes in the presence of CHCl$_3$ as well as influences of acetate and halide ions on facet evolution during secondary growth steps have been reported. [15-19]

Halides or halogen compounds have proven to severely influence the ligand sphere and chemical reactivity of NPs and even to support the attachment of CdSe NPs to carbon surfaces.[20] There are examples of different materials, such as MnS, CoO and CdSe where hexagonal bullets, pyramidal, pencil or diamond shapes are formed when chloride precursors are employed, which points to a general effect of chloride in wurtzite structures.[18,21,22] Apart from influencing NP shape, atomic X-type (anionic) ligands released from halogen compounds such as molecular chlorine, halide salts and alkylsilyl chlorides were shown to improve the optical properties of NPs and/or to increase the electrical transport in NP thin film or solar cell devices by replacing long alkyl and insulating ligands.[23] Furthermore, by tuning the exposed crystallographic facets, size and reactivity of the NPs numerous possibilities for self-assembly and the deposition of other materials in the preparation of hybrid NPs are achieved.[24,25] For this reason, it is interesting to systematically examine the influence of different halogen compounds on the shape evolution of NPs.

In a hot-injection synthesis, NPs undergo different stages from nucleation over growth to ripening, which are closely related to the supersaturation of the reaction solution and its depletion. In the following, we provide new insights into the formation of rod-shaped NPs and their morphological evolution to hexagonal pyramidal CdSe NPs induced by halide ions which are released in-situ from different halogen compounds. Depending on the type and chemical structure of the halogen molecule chosen, growth kinetics as well as facet passivation vary, which strongly influences the shape evolution at different stages of the reaction. From the here reported studies we can rationalize that the release of halide ions and their availability during nucleation and/or growth play a major role in the shape evolution. Synchrotron XPS measurements and theoretical calculations examining the ligand sphere are presented to explain why pyramidal shapes evolve under the influence of halogen compounds. Our studies contribute to the general understanding of the kinetics of shape evolution and expand the knowledge on in situ manipulation of ligands and thus nanoparticle surfaces. With this methodology it is possible to extend the number of feasible nanoparticle geometries and to design the shape of NPs in regard to their future application, for instance in solar cell devices where NP size and shapes of CdSe sensitizers have an influence on the charge injection into TiO$_2$ electrodes.[26]

## 2. Experimental section

**2.1 Materials.** Cadmium oxide (CdO; 99.99+%) was purchased from ChemPur. Tri-n-octylphosphine (TOP; ≥90% and distilled, stored in a nitrogen filled glovebox), 1,2-dichlorobutane (1,2-DCB; 98%), 1,1,2-trichloroethane (TCE; 96%), 1-chlorooctadecane (COD; 96%), 1,2-diiodoethane (DIE; 99%), dodecyltrimethylammonium bromide (DTAC; >99%) and selenium shots (amorphous, 2-4 mm, 0.08-0.16 in, 99.999+%, stored in a nitrogen filled glovebox) were ordered from Aldrich, while 1,2-dichloroethane (DCE; p.A.), tri-n-octylphosphine oxide (TOPO; >98%), Acilit pH paper, methanol and toluene were obtained from Merck. Acros is the producer of our 1,2-dibromoethane (DBE; 99%) and 2,3-dichlorobutane (2,3-DCB; 98%). Octadecylphosphonic acid (ODPA, 97-98%) was bought from Alfa Aesar. Highly oriented pyrolytic graphite (HOPG) substrates of ZYB quality (1 cm2/ 2 mm) were purchased from NT-MDT. All chemicals except TOP were used without further purification.

**2.2 Methods.** *Hexagonal CdSe nanopyramids* were prepared following a previously published method with minor modifications.[25] In a three necked flask equipped with a condenser, a septum and a thermocouple in a glass mantle, 25 mg (0.19 mmol) CdO, 0.14 g (0.42 mmol) n-octadecylphosponic acid (ODPA) and 3.0 g (7.8 mmol) tri-n-octylphosphine oxide (TOPO) were heated to 120 °C for 30 minutes. During this time, two switches from vacuum to nitrogen and back were carried out. The temperature was raised to 270-290 °C for complexation until an optically clear and colorless solution was obtained (around 70 minutes). After cooling down to 80 °C, 10 µL (0.13 mmol) of dichloroethane (DCE) were injected and the temperature was raised again. At 265 °C 0.42 mL (0.42 mmol Se) of a 1 M Se in TOP solution were injected before reducing the temperature to 255 ± 2 °C for growth. The color of the solution turned from colorless via yellow and orange to brown, indicating the formation of CdSe NPs. The reaction was quenched after 4 h by cooling down to 70 °C and injecting 3.5 mL of toluene. The resulting CdSe pyramids and all samples taken during the reaction were purified by three cycles of precipitation with methanol (addition until turbidity occurs), centrifugation (4500 rpm, 3 minutes), removal of the supernatant and re-dispersion in toluene for further characterization.

When halogen sources different from DCE were employed, they were added 10 °C below their boiling point, at 80 °C dodecyltrimethylammonium chloride (DTAC) or in the case of 1-chlorooctadecane (COD) directly at 265 °C. Amounts and temperature values at the moment of addition are listed in Table S1 in the Supporting Information. Warning on the use of 1,2-diiodoethane in the reaction: The strong acidity of HI that can be formed in the reaction with DIE leads to production of considerable (smelly and thus dangerous!!) amounts of H$_2$Se, which can be carried out of the apparatus with the inert gas flow when conducted in a laboratory hood.

*Separation of rod formation and morphological evolution to pyramids.* In the case of a post-nucleation addition of the chlorine compound, the same procedure as described above was followed except that 1-chlorooctadecane (COD) instead of DCE was used and injected 15 minutes after the injection of Se in TOP at the growth temperature of 255 ± 2 °C. The growth period was maintained for 24 h and several aliquots were taken.

*The pH of aliquots* was determined by extraction with water. Aliquots of 0.15 mL were taken at different stages of the reaction and diluted with 0.5 mL toluene. From this, 300 µL were vortexed with 150 µL of water, shaken for 15 minutes and left to settle for 1 hour. To improve phase separation, the samples were centrifuged for 2 minutes at 1000 rpm. Drops of the aqueous layer were tested with Acilit pH paper.

**2.3 Characterization.** Absorption and emission measurements were carried out in quartz vessels with an optical path length of 10 mm. UV-vis absorption spectra were obtained with a Varian Cary 50 Spectrophotometer (one-beam), while emission spectra were obtained with a Horiba Jobin Yvon Fluoromax-4 spectrophotometer. TEM images were obtained with a JEOL Jem-1011 instrument at an acceleration voltage of 100 kV. HR-TEM images stem from a Philips CM 300 microscope with an acceleration voltage of 200 kV. Samples were prepared by drop-casting 10 µL of the purified sample onto a copper grid covered with amorphous carbon. Nanoparticle dimensions were determined as medium values of 210 particles measured with *Image J* software.

Samples for XPS were prepared by adding a 0.5x0.5 cm² piece of freshly peeled highly oriented pyrolitic graphite (HOPG) to the solid compounds before complexation of CdO, following previously published methods.[20,27] The reaction time was prolonged to 21 h as a deceleration of the transformation occurs with carbon substrates inside the reaction pot and optimum attachment takes approximately this time. In XPS measurements the samples were excited with 620 eV synchrotron radiation in the SurICat UHV XPS set-up connected to the BESSY II storage ring, Helmholtz Zentrum Berlin. The ratios between different surface species have been estimated as the ratios between the corresponding intensities weighed by the corresponding sensitivity factors. *Note:* When comparing bulk and surface species, the interpretation of the ratios is not straightforward. This is due, among other reasons, to the fact that photoelectrons arriving at the detector arise mostly from the region of the sample close to the surface and the measured intensity of the chemical species sitting far away from the surface decreases exponentially with the distance from the surface. Moreover, the constant for the exponential decay is not generally known. As a result, the observed intensities of the bulk elements might not reflect the total amount of bulk material present in our samples, especially since the diameter of the NPs is already larger than the average mean-free path for the photoexcited electrons. Such ratios can thus be interpreted only in a semi-quantitative fashion.

For *Total reflection X-ray fluorescence spectrometry* (TXRF) measurements with a Bruker Picofox S2 machine the samples were purified by several precipitation and re-dispersion cycles. From a solution of 0.2 mL of aliquot and 0.5 mL of toluene 0.3 mL were precipitated by addition of methanol (1:1) and centrifugation (3 min/4500 rpm), re-dispersed in hexanes and centrifuged again (2 min/4500 rpm). The supernatant was mixed with methanol/ethanol (1:1:1), centrifuged (3 min/4500 rpm) and re-dispersed in toluene before methanol was added (1:2) and the sample centrifuged again (3 min/ 15000 rpm). The sample after 10 minutes was purified further (hexanes/ethanol 1:1, 3 min/ 4500 rpm, toluene/methanol 1:1, 3 min/4500 rpm) due to the high amount of remaining Cd-ODPA precursor (visible as slimy precipitate). For measuring the samples were dispersed in toluene and drop-casted onto a silicon dioxide substrate.

# 3. Results and discussion

**3.1. Shape transformation of CdSe rods by dichloroethane** In earlier studies we observed a morphological transformation of hot-injection prepared CdSe nanorods to hexagonal dipyramidal particles (CdSe pyramids) during the high temperature preparation of composites with carbon allotropes dispersed in dichloroethane (DCE).[27,28] Similar NPs in terms of size and shape (although more polydisperse) can be obtained by adding HCl to the nanorods-CNT reaction.[27] To learn more about the occurring shape evolution, we studied the reaction in a modified protocol without carbon allotropes but with varied halogen compounds. CdSe pyramids were prepared after a previously reported protocol, with modifications.[25] In a typical synthesis, 10 µL (0.13 mmol) of DCE were added to a complex of cadmium with octadecylphosphonic acid (ODPA) in trioctylphosphine oxide (TOPO) at 80 °C and the mixture reacted with selenium in TOP (1 M, injected at 265 °C) at 255 °C for four hours (Cd/Se/ODPA/DCE = 1:2:2:0.7).

In an experiment without DCE, the shape evolution follows the reported mechanism where kinetic growth control causes the NPs to grow along the c-axis of the wurtzite structure, resulting in rod-shaped NPs shortly after the Se-injection.[2,29,30] During the course of the reaction, the monomer concentration depletes and the NPs grow in all directions before thermodynamically controlled ripening results in equilibrium shapes with aspect ratios approaching one and Ostwald ripening eventually increases the polydispersity of the sample.[2,29]

As a comparison, in the presence of DCE the NP size increased in the kinetic growth regime as indicated by a shift in the absorption maxima in Figure 1a. Furthermore, while in the absence of DCE the rods maintain the elongated shape after 4 h (for TEM images of control samples see Figure S1, in the supporting information) thermodynamically controlled ripening lead to hemimorphic hexagonal dipyramids with a wurtzite crystal structure in the presence of DCE, as it can be seen in Figure 1 (XRD in Figure S2).

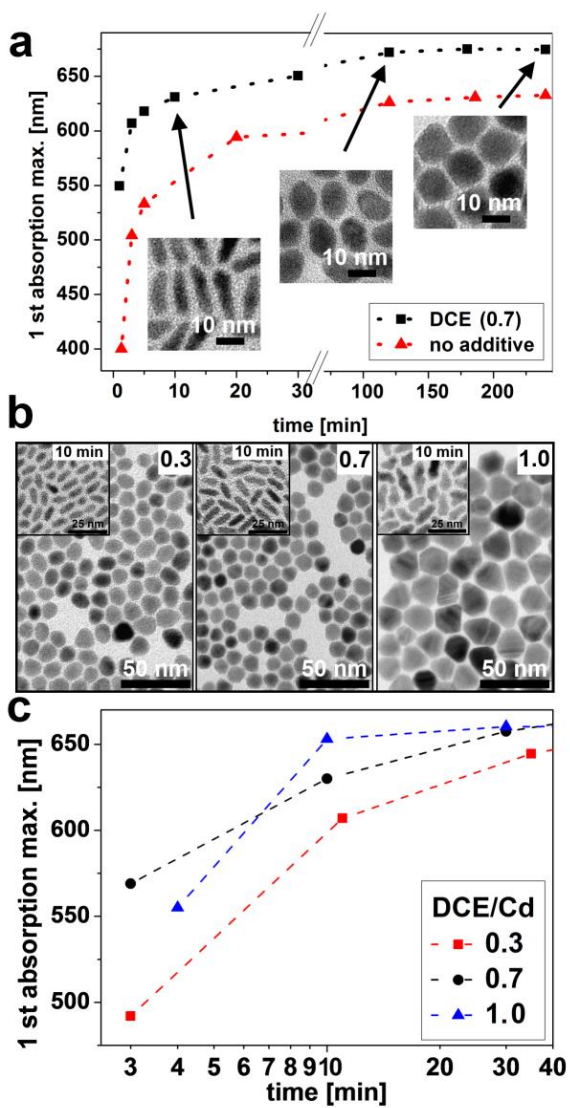

Figure 1. a) Comparison of the evolution of the first absorption maxima of aliquots with and without DCE (DCE/Cd: 0.7). A shift of the absorption maxima to higher wavelengths with DCE and smaller shifts between later samples indicate the formation of bigger NPs and an earlier transition to ripening processes. TEM images show aliquots from a synthesis with DCE after 10, 120 and 240 minutes. b) TEM images of samples with different amounts of DCE (DCE/Cd top right inset) after 10 min and four hours. (c) An increase of the DCE/Cd ratio from 0.3 to 0.7 results in a smaller band gap in early reaction stages and a flatter slope between the absorption maxima with time. A ratio of 1.0 leads to an inhibition of particle formation before a fast shift occurs which then levels off quickly when ripening sets in.

A variation of the amount of DCE added from a DCE/Cd ratio of 0.3 to 1.0 revealed two tendencies occurring with increasing amount of DCE: 1) the size of rods after (e.g.10 minutes) increased with higher amount of DCE; 2) bigger rods evolved into bigger nanopyramids and the diagonal facets were more pronounced (see TEM images and UV-vis absorption spectra in Figure 1b and insets; NP dimensions are DCE/Cd 0.3: 10.4 ± 1.1 x 4.4 ± 0.4 nm (10 min); 12.8 ± 2.0 x 8.3± 1.0 nm (4 h); 0.7: 13.0 ± 1.3 x 4.7 ± 0.5 nm (10 min), 13.1 ± 1.4 x 12.2 ± 1.3 nm (4 h); 1.0: 16.8 ± 2.6 nm x 7.6 ± 0.8 (10 min), 20.8 ±2.3 x 20.9 ± 2.4 nm (4 h)). In UV-vis spectra, the mainly kinetically driven rod growth is characterized by a rapid red-shift of the absorption maxima occurring during approximately the first 30 minutes of the reaction (Figure 1a, c). Minor shifts are observed afterwards, indicating the formation of thermodynamically more favorable shapes through ripening caused by precursor/monomer depletion.[31] For a DCE/Cd ratio of 0.3 and 0.7 growth lasted about 30 minutes and, at a ratio of 1.0 an inhibition of particle formation (no nucleation visible after 3 minutes) was observed before a fast shift occurred which then leveled off after 10 minutes when ripening set in (Figure 1c). The slope for DCE/Cd 0.7 is also smaller than for 0.3, which can be interpreted as a faster transition to the ripening stage with increasing concentration of the chlorine compounds. Regarding the homogeneity in size and shape, the ratio of 0.7 seems to be high enough to promote faceting of all NPs while it is low enough to prevent the formation of irregular rods and less defined pyramids. The formation of rough side facets instead of flat {10-10} ones in rods grown with a DCE/Cd ratio of 1.0 can be related to preferential formation of Cd-rich sites and will be discussed later on. The irregular shape, size, and size distribution of the initial rods influences the quality of the later formed pyramids in terms of stacking faults, and inhomogeneities in shape and size distribution. A DCE/Cd ratio of 1.3 led to the same outcome as when $CdCl_2$ (without DCE) was used as Cd precursor instead of CdO: no colloidal solution but bulk precipitate was formed (see XRD in Figure S3). At a DCE/Cd ratio of 6.5, no nucleation occurred at all, which resembles earlier attempts of using $CdCl_2$ or salts of other strong acids as precursors where the authors reported a high solubility of the respective salts in the reaction mixture and an inhibition of NP nucleation.[32,33]

To better understand the mechanism behind the shape control and how it eventually can be utilized to control particle geometries, we performed a series of experiments with different halogen compounds as presented in the following.

**3.2 Effects of the chemical structure of the halogen compound** If the additive affects the kinetics of rod formation and ripening, it should be possible to employ structural properties of the initial halogen compounds to manipulate the progression of the reaction. For this reason, we carried out reactions with different chloroalkanes, namely: 1,2-dichloroethane (DCE), 1,1,2-trichloroethane (TCE), 1,2-dichlorobutane (1,2-DCB), 2,3-dichlorobutane (2,3-DCB) and 1-chlorooctadecane (COD). All additives were injected 10 °C below their boiling point or, as in the case of COD with a boiling point of 348 °C, directly at 265 °C. The reactions were followed by absorption spectroscopy and TEM.

In all cases, pyramidal NPs were obtained but the temporal evolution and homogeneity of the samples differ notably. Comparing the evolution of the absorption maxima of the NPs upon addition of the same concentration of different chloroalkanes (Figure 2), indeed suggests a correlation between the molecular structure of the additive and the rate of the evolution of the NPs. The reactivity of the employed chloroalkanes follows the tendency expected from their structural properties in a substitution reaction: The less the chlorine atoms are stabilized by the neighboring alkane groups (+I effect and steric stabilization), the higher are the absorption maxima in early reaction stages and the bigger are the corresponding nanorods (less affected growth; for TEM see Figure S4). Regarding the shape evolution during ripening, pyramidal NPs were eventually obtained in all cases but the slope of the absorption maxima in Figure 2 becomes shallower with decreasing steric hindrance of the molecule which indicates a faster transition from rod growth to ripening (TEM images and extended absorption data in Figure S4).

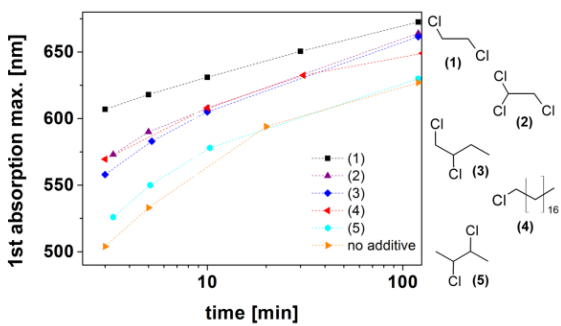

Figure 2. Evolution of first absorption maxima with different chloroalkanes (1-5) compared to the evolution without additive. The graphs correspond to (1) 1,2-dichloroethane (DCE), (2) 1,1,2-trichloroethane (TCE), (3) 1,2-dichlorobutane (1,2-DCB), (4) 1-chlorooctadecane (COD) and (5) 2,3-dichlorobutane (2,3-DCB) samples. Molecules with sterically less stabilized chlorine atoms lead to bigger NPs (longer absorption wavelengths after 3 min) and a faster shape evolution to faceted pyramids (more shallow slope of absorption maxima).

If the boiling point of the haloalkane is higher than the NPs growth temperature (255 °C) it can be added at a time well separated from nucleation and initial rod growth. In this way, the nucleation and first stages of growth are not affected by the chloro compound which allows synthesizing smaller nanopyramids with narrow size distribution. This is the case of adding 1-chlorooctadecane to already formed, comparatively small and homogenous rods (c-axis: 5.6 ± 0.7 nm) directly at the growth temperature, where smaller hexagonal pyramids (c-axis length of 7.8 ± 0.7 nm) were obtained (Figure S5). This shows that rod growth and pyramid formation can be separated and the length of the c-axis in both geometries is related to each other, with smaller rods allowing for the preparation of smaller pyramids.

**3.3 Increased rod size and reduced tendency towards pyramid formation with heavier halogens** To investigate the possibility to further influence the kinetics of the reaction we employed chemical analogs of 1,2-dichloroethane (DCE), namely 1,2-dibromoethane (DBE) and 1,2-diiodoethane (DIE) in equimolar amounts (DXE/Cd 0.7). In the kinetic growth regime, anisotropic NPs terminated by zigzag shaped side facets that are even larger (>20 nm in c-axis) than observed in the case of a DCE/Cd ratio of 1.0 evolve with both, DBE and DIE. After ten minutes (TEM: Figure 3a-c) the DIE samples are the largest, which hints at a stronger influence of iodoalkane followed by the bromoalkane during nucleation and early stages of the reaction compared to chloroalkanes. This trend suggests that the morphological effects on the NPs must be related to the reactivity of the haloalkanes, with increasingly nucleophilic halogen atoms making a bigger impact. In the ripening stage, NPs prepared in the presence of DBE exhibit a clear tendency towards pyramid formation, while NPs prepared with DIE exhibit smoothed surfaces but show only a slight tendency of growth perpendicular to the c-axis (Figure 3e-f, h-i). From this we can derive that different haloalkanes can influence the NP shape to different extends in kinetic growth and ripening regimes.

**3.4 The active species: halide ions** Due to the previously observed effects, where a higher amount of additive as well as sterically more accessible and more nucleophilic halogen atoms in the molecule result in bigger nanorods and more strongly faceted pyramids, the active species influencing both nucleation and growth seem to be released halide ions (chloride, bromide and iodide). Further evidences of the anion rather than the haloalkane molecule affecting the NP growth were obtained by adding HCl as chlorinated compound or a salt, such as dodecyltrimethylammonium chloride.[27] In both cases, pyramidal shapes were obtained even though the ammonium salt produced more bullet-shaped morphologies (see reference 32 and Figure S6). Our findings fit with other results showing that HCl (aq) can induce structural changes in Cd-chalcogenide NPs and that chloride ions exhibit a tendency to induce pyramidal or diamond-like shapes in wurtzite materials.[18,21,34,35] All of this suggests that chloride rather than DCE as a molecule is responsible for the observed effects and that chloride can be added to the reaction in the form of a salt, a mineral acid or haloalkanes. The release mechanism is related to a reaction between the haloalkane and TOP to a $(R_4P^+(halide)^-)$-type compound as suggested by Lim et al.[19] and investigated in detail by Palencia et al. (submitted)[36]

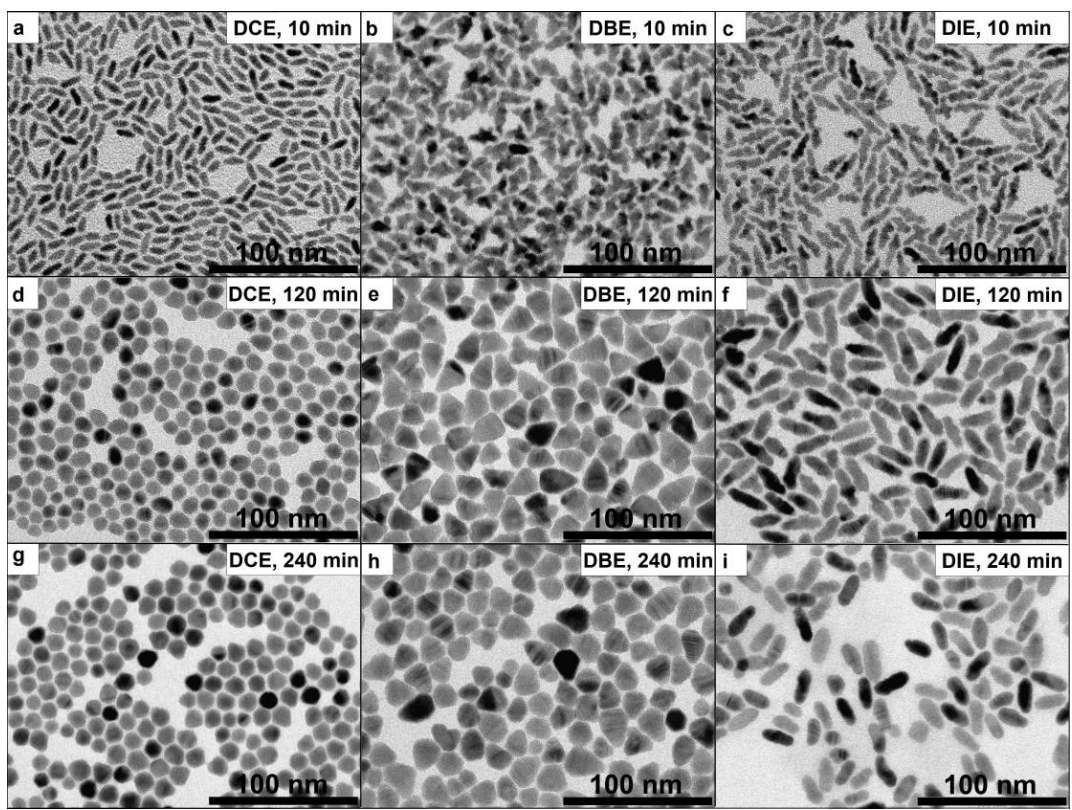

Figure 3. TEM images of reactions with DCE, DBE and DIE (DXE/Cd 0.7). (a-c), 2 h (d-f) and 4 h (g-i) DBE and DIE increase the size and surface roughness of rod in samples extracted after 10 min (a-c). With time, ripening smoothens the surface facets (2 h, e and f) but after four hours (g-i), CdSe NPs with different di-haloalkane additives remain at different stages of ripening and show significant differences in dimensions.

**3.5 Incorporation of halides into the ligand sphere**

In order to verify changes in the ligand shell produced by the different haloalkanes, synchrotron XPS data of NPs prepared without additive (rods) and pyramidal NPs prepared with DCE (Cl-pyramids) and DBE (Br-pyramids) have been compared in Figure 4. NPs produced in the presence of DCE show a clear Cl peak (Figure 4a) centered at 198.8 eV, an intermediate binding energy between the position of the Cl $2p_{3/2}$ in $CdCl_2$ (198.4 eV) and that of Cl $2p_{3/2}$ in the more ionic $ZnCl_2$ (199.1 eV). This position might be caused by an ionic interaction between chloride and the surface of nanopyramids. The presence of the Br 3d signal in the NPs produced with DBE (Figure 4b) suggests that Br from DBE reacts similarly to Cl. Both Cl and Br signals turn out to be comparatively broad (FWHM Cl: 1.2 eV, Br: 1.3 eV) with respect to the Au signal reference (0.7 eV for the FWHM of Au 4f peak) which indicates that these elements can bind indistinctly to several facets. From the XPS spectra of Figure 4c, quantitative analysis of the surface-sitting elements P and Cl (or Br) shows that, for Cl-pyramids the ligand shell is composed by Cl species by about (43 ± 3)%, and the ligand shell of the Br-pyramids contains (31 ± 5)% of Br. TEM images in Figure 4c show that the NPs with Br are less sharply faceted than with Cl at the same reaction time. Together with the previously shown trend (more DCE leads to more sharply faceted NPs) this speaks for a correlation between the particle shape and the amount of halogen in the ligand sphere. In previous work we ascertained a correlation between P/Cd and Cl/Cd peak area ratios and observed that the incorporation of Cl into the ligand sphere of CdSe NPs prepared in presence of DCE is accompanied by a change of the P signal. Solid NMR, ICP-MS and XPS showed a difference in the ligand sphere with rods containing a mixture of $ODPA^{2-}$ and anhydrides and pyramidal NPs containing less P with mostly $ODPA^{2-}$ and $Cl^-$ ligands on the surface.[20] Regarding the P content of the samples shown in Figure 4, the sample without additive shows a ratio between the area of the peaks of P and Cd of 0.14, 0.09 is observed for Br-pyramids and 0.05 for Cl-pyramids. This confirms that samples with smaller P content (relative to Cd) show an increased halogen content.

The presence of both halogen and P atoms in NPs with DCE was also confirmed by total reflection X-ray fluorescence (TXRF) measurements of purified samples (for values see Table S2 in SI). An increase in the Cl/Cd and a decrease in the P/Cd ratio were observed with proceeding reaction times, which supports the idea of a gradual incorporation of $Cl^-$ into the ligand sphere.

The above findings suggest a relation between the amount of phosphonic acid ligands, their ratio to halide and the final nanoparticle morphology and can be understood as dynamic changes in the ligand sphere during ripening.

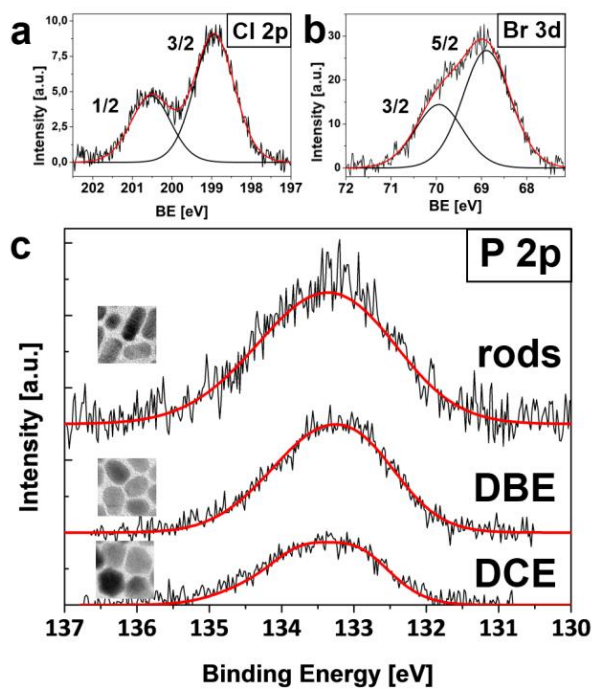

Figure 4. Synchrotron XPS data of the regions of Cl 2p (peak position 3/2: 198.8 eV, (a)) and Br 3d (peak position 5/2: 68.9 eV, (b)) binding energies obtained at a photon energy of 620 eV. The areas of the P signals in (c) are normalized to the respective Cd 3d signal to derive the atomic ratios. The ratio of the area between the P 2p and Cd signals (not shown) decreases from rods (0.14) to NPs produced with DBE (0.09) and DCE (0.05).

**3.6 Ligand adsorption energies favor pyramidal CdSe shapes** Ligand adsorption plays a crucial role in the growth of NPs because they influence the relative growth rate of the facets or even determine which type of ion dominates the surface.[37,38] We performed basic density functional theory (DFT) calculations to examine the binding affinities of ligands and additives to the dominating facets of hexagonal pyramids by employing the ORCA software.[39] The LDA exchange functional, the correlation functional VWN-5[40] and the Ahlrichs TZV basis set[41] served as framework for the calculations. Adsorption energies were determined by comparing the sum of the separate energies of CdSe crystal and ligand molecule with the total energy of the complete system. During simulation under aperiodic boundary conditions, the CdSe crystals were fixed to the experimental lattice constant and the geometry of a pristine facet while the ligand molecule was free to relax (the dimensions of the slab are visible in Figure 5). In order to keep the simulation time short, instead of simulating the whole ligand molecule, we used shorter versions with the functional group and a three carbon aliphatic chain (e.g. instead of octadecylphosphonic acid we simulated propylphosphonic acid, PPA). This does not change the qualitative aspects of the results.

Our calculations (Table 1) state that the higher charge a ligand species carries (X-type-ligands: phosphonate species, halides), the stronger the adsorption is, while neutral ligands (L-type ligands: TOP, TOPO, protonated phosphonic acid species, DCE) interact only weakly on all facets, in accordance with previous reports.[42-44] All species adsorb comparatively weakly to the Se-rich (000-1) facet, while they adsorb more strongly to Cd-rich sites.[37,42,45]

Table 1. Adsorption energies of different ligand molecules (PPA: Propylphosphonic acid, TPP(O): Tri-n-propylphosphin(e oxide)) on distinct CdSe facets calculated by the DFT method.

| Ligand | Adsorption energies [eV] | | | |
|---|---|---|---|---|
| | CdSe (0001) | SeCd (000-1) | Side (10-10) | Slope (10-1-1) |
| PPA | 1.86 | 0.97 | 2.28 | 2.15 |
| PPA-anhydride | 2.18 | 0.97 | 2.82 | (6.23)[a] |
| TPP | 1.76 | 1.04 | 2.08 | 1.81 |
| TPPO | 1.93 | 0.91 | 2.37 | 2.22 |
| DCE | 0.90 | 0.49 | 1.19 | 0.77 |
| PPA2- | 11.44 | 6.11 | 9.03 | 12.90 |
| PPA-anhydride2- | 10.40 | 4.40 | 8.26 | 12.57 |
| Cl- | 4.39 | 1.80 | 3.45 | 5.49 |
| Br- | 4.14 | 1.70 | 3.21 | 5.20 |
| I- | 4.07 | 1.80 | 3.15 | 5.06 |

[a] Decomposes on this facet.

For the neutral ligands ODPA, TOP, and TOPO the adsorption energy to the different facets decreases in the following order: (10-10) side > (10-1-1) slope > (0001) bottom > (000-1) top, meaning that the ligands bind strongest to the side facet, which confirms earlier calculations explaining the rod growth,[42] and closely followed by the sloped facet. However, the binding affinity is weak compared to the charged ligands.

For the charged ligands ODPA2- and ODPA-anhydride$^{2-}$, and also for halogen ions the adsorption energy decreases in the following order: (10-1-1) sloped > (0001) bottom > (10-10) side > (000-1) top, meaning that the ligands bind strongest to the sloped facet, closely followed by the Cd-rich bottom facet. Their adsorption energy is much higher than the one for neutral molecules. They most strongly interact with the sloped (10-1-1) facet, since this is the roughest one exposing Cd sites. In Figure 5 it can be seen that on the (0001) facet (bottom) Cd is coordinated by 3 Se atoms, compactly. On the (10-10) facet (side) Cd is also coordinated by 3 Se atoms, but in a much more corrugated fashion. Finally, on the (10-1-1) facet (slope) the Cd atoms are only coordinated by 2 Se atoms and the surface unit cell is larger. As a result the charged ligands, especially the important ligand ODPA$^{2-}$, have the largest adsorption energy with the rough, sloped facet. Neutral phosphonic acid anhydrides resulting from protonation (see discussion below) were instable on the (10-1-1) sloped facet and even decomposed under the same calculation conditions, while the halide ions stabilize on the same facet. The halide ions clearly follow the trend of electronegativity and the bonding strength decreases in the following order: Cl- > Br- > I-.

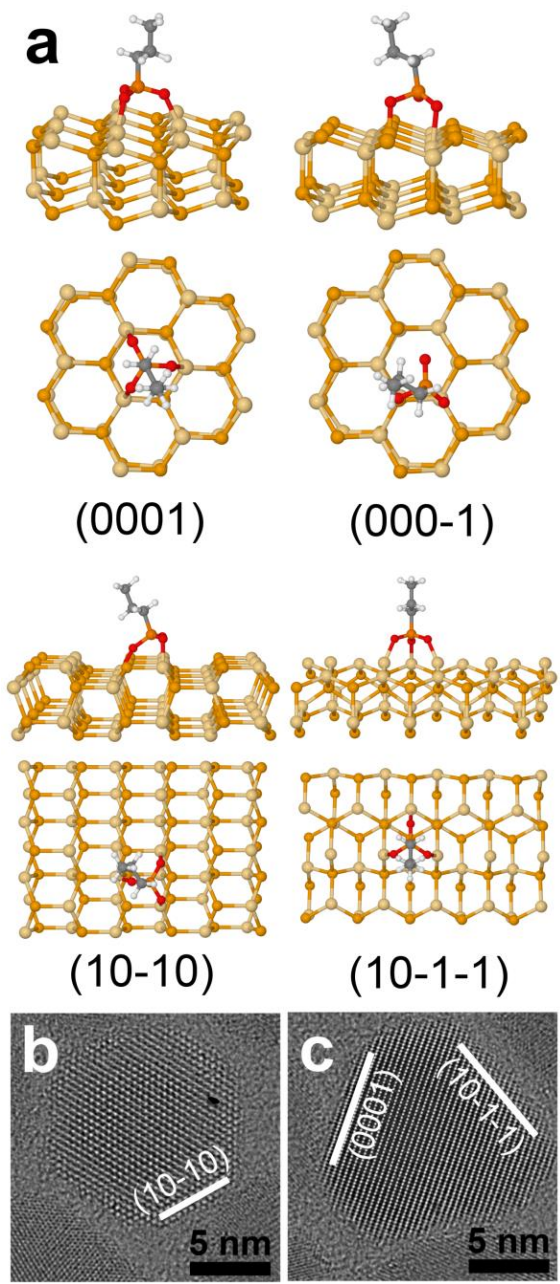

Figure 5. (a) Side and top-view of PPA2- adsorbed to different CdSe facets and (b) HR-TEM images of pyramidal CdSe NPs as (b) top- and (c) side-view with indicated facets. Beige: Cd atoms; orange: Se atoms.

The halides as singly charged species possess an adsorption energy between the neutral ligands and the double-deprotonated ODPA species.

From the results given in Table 1 it is clear that halide ions only are able to displace TOP binding to Se sites or ODPA and anhydrides in a neutral form (L-type coordination). Mixed coordination with Cl- and ODPA$^{2-}$ on Cd-rich facets can be explained as energetically favorable as the steric demand of deprotonated phosphonic acid molecules leaves room in between them for the adsorption of small halides. Such mixed coordination is in agreement with reports on trap filling by halides in PbS samples, where halides "squeeze in" between or even partially substitute oleic acid ligands.[23,46]

**3.7 Influence of halide ions on the shape evolution**
By comparing all results it becomes clear that there are different effects of the halides in the kinetic growth stage and the thermodynamic growth stage and differences in the degree to which the compounds affect the NP morphology. Based on the calculated adsorption energies and the experimental findings, a picture of the shape evolution process can be drawn. Figure 6 graphically correlates the process with the reaction stages.

**3.7.1 Rod growth** In the kinetic growth regime at high monomer concentrations (000-1) Se facets, which are least strongly capped with ligands, exhibit the highest growth rate and rods with strong side-facet protection by ODPA2- and deprotonated anhydrides are formed. This was observed also in the presence of any of the here employed halogenated compounds. Up to a molar DCE/Cd ratio of 0.7 and with all other chlorine compounds added in equimolar amounts, the rods formed in the early stage of growth become bigger. These observations together with prior results indicating that DCE modifies the Cd-ODPA complex,[27] suggest that DCE/Cl- increases the solubility of CdSe monomers and/or the Cd-precursor, for instance by forming mixed coordinated complexes as reported for Cd-amine complexes with chlorine ligands.[34,47] An increase of the monomer solubility leads to an earlier transition from nucleation to growth by increasing the critical supersaturation for nucleation and results in larger NPs.[48] Similar trends were reported for other Cd-chalcogenide/ligand systems when increasing amounts of halides were present in the reaction and can be explained by enhanced stabilization of precursor complexes and monomers by halogen ligands.[18,19,34]

Above a DCE/Cd ratio of 0.7 and with DBE and DIE zigzag faceting parallel to the c-axis is observed. Similar results were obtained in experiments where hexylphosphonic acid was employed to promote growth of higher aspect ratio rods: Above a certain threshold zigzag structured surfaces and arrow shapes are formed.[49] The latter suggests the formation of Cd-rich facets, which would mean that HPA or halogen compounds compete with ODPA for surface coordination and above a threshold the shorter ligand induces the expression of Cd-rich diagonal facets already during anisotropic growth. If nucleation and growth are affected by the presence of halide ions competing with ODPA, the nucleophilicity of the elements should be taken into account. The nucleophilicity increases in the line Cl- < Br- <I-. Both Br- and I- are thus released more easily from haloalkanes than Cl- and should be available in higher amounts at earlier stages, so they can disturb the evolution of the side facets to a higher extend, as observed in Figure 3. By means of TXRF we ascertained that the halogen/Cd ratio of samples produced at similar reaction times (30 min) was higher for DIE than for DBE and DCE (see Table S3).

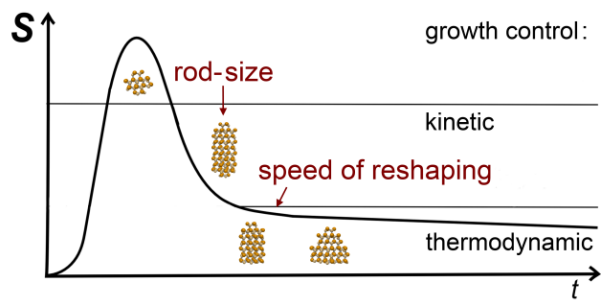

Figure 6. Schematic LaMer plot with different stages of pyramid evolution. The shape evolution begins when the saturation S reaches the level for thermodynamic growth control or the ripening regime (developed from 50). With halogen compounds both the size of first formed rods and the rate of ripening can be controlled in regard to the molecular structure of the compound.

**3.7.2 Shape evolution in the ripening stage** In the ripening stage, crystals approach equilibrium shapes to minimize their surface energy. Rods are expected to finally evolve into NPs with an aspect ratio of one.[29] With additional halogen compounds, the aspect ratio also tends towards 1 with proceeding reaction time but the equilibrium shape is changed to a polyhedral morphology. Noteworthy, pyramidal shapes are also found in cadmoselite minerals in nature.[51] In the particular case of the ripening in the presence of chloride the weak passivation of the (000-1) facet may trigger the initial etching promoting further ripening, which agrees with the fact that rods evolve to the pyramidal shape exclusively from one of the tips in the presence of DCE.[52] By TEM tomography studies we ascertained that this tip is, in fact, the (000-1) facet of CdSe.

However, although a direct etching of Se by chloride is feasible,[53] the presence of water in the reaction may also play a critical role during the ripening stage. This water, a byproduct of the Cd complexation with ODPA,[54] may (i) increase the conversion of TOPSe,[55] (ii) provide protons to activate Cd-ODPA complexes,[56] as well as (iii) help to desorb Se2- and ODPA from the surface by protonation (neutral ODPA binds less strongly than ODPA$^{2-}$, see Table 1) and (iv) lead to a higher polarity of the reaction mixture (TOP/TOPO). We examined the relative amount of free protons in the present reaction by extracting different aliquots with water and observed that the pH was acidic in all samples and decreased with time (see Experimental section for further details). A significant step down occurred from 10 to 30 minutes, just when the rods began to ripen in the presence of DCE. This means that the presence of water also plays an important role during ripening and in fact, more faceted pyramidal NPs are obtained in its presence than when it is removed *in vacuo* before Se is injected (not shown).

To understand the ripening process, once the Se atoms begin to be removed from the (000-1) facet, halide ions are able to occupy surface sites of forming facets and eventually hinder the attachment of sterically demanding ODPA-anhydrides or ODPA$^{2-}$. The diffusion of the comparatively small anions might be favored by the presence of water which is able to protonate the tightly bound ODPA2- and anhydride decreasing their binding energy (see Table 1) thus, allowing for displacement of ODPA species. Alternatively also inaccessible Cd sites for the bulky ODPA species can be passivated by chloride. The small Cl- allows for a fast diffusion (faster than the other halides) providing a stronger advantage versus phosphonate ligands in the competition for newly formed Cd surface sites. Furthermore, from data shown in Table 1, ODPA anhydrides are instable on the sloped (10-1-1) facet, while all halide ions show relatively high adsorption energy values for this facet. Thus, based on our results, we believe that the co-adsorption of phosphonates and Cl on the sloped facet is the key to the obtained pyramidal NPs.

NPs show less pyramidal shape in the presence of DBE and rod-shape in the case of the DIE. To understand this effect we should take into account that iodide and bromide ions showed a more prominent role during nucleation and first stages of growth (as a result of a higher reactivity and thus, higher relative concentration, as ascertained by TXRF measurements). During the ripening regime their effect was attenuated compared to the case of chloride, as shown in Figure 3 g-i. Thus, the slightly lower adsorption energy of I- < Br- < Cl- or, in other words, the stronger Cd-Cl bond compared to Cd-Br and Cd-I must also have an influence in the final form during the ripening stage.

The shape evolution presented here takes place in a reaction solution with halogen precursor molecules where dynamic absorption/desorption processes of ligands and surface atoms facilitate the reorganization of the particle surfaces to minimize the surface energy. Examples of pyramidal or bullet-like morphology have been reported caused by a modification in the chemical composition of the ligand sphere by a short and labile ligand in combination with monomer addition,[17] or by etching of Se-sites of spherical CdSe NPs in heated, basic solutions containing a strong Cd2+ ligand.[15] We believe that the shape transformation from rods to pyramids in our case is similar to an acidic etching process with hydrogen halides, combined with the growth of facets stable in their presence. The hexagonal pyramidal shape with a positively charged base (0001) and cationic diagonal {10-1-1} facets was indeed reported as the equilibrium form of CdSe, CdS and other II-VI compounds etched by HCl.[38,57]

## 4. Conclusions

Halogen compounds severely impact the shape evolution in the hot-injection synthesis of CdSe nanorods. In the presented method haloalkanes influence both the kinetic and thermodynamic growth regime of the reaction leading to hexagonal pyramidal equilibrium shapes which form through ripening of rod-shaped intermediates.

We found that such additives work very similar to ionic halogen additives and that structural properties and concentration of haloalkanes determine the rate of the morphological evolution. The tendency to release halide ions was found critical for the influence on the shape, with comparatively more halides present in the kinetic growth regime leading to larger rods which, above a threshold, exhibit zigzag facets parallel to the c-axis. Larger rods evolve into larger pyramidal NPs. A variation of the halogen compound showed that with I- the effect on the kinetic growth is stronger than for the reshaping to pyramids while this was the opposite for Cl- and intermediate for Br-. In this way, the kinetics of in situ X-type ligand generation can directly be exploited to influence the final particle structure.

Synchrotron XPS and TXRF data showed that the evolution is accompanied by a loss of phosphonate ligands and incorporation of atomic halide ligands into the ligand sphere. We expect this method to be transferable to other systems which can be shape-tuned by halides such as other II-VI semiconductors and thus become a versatile tool for the design of new particle shapes. Mixed coordination of long chained ligands and halides is expected to proof beneficial for electrical conductivity.[58] A thus decreased distance between NPs in devices was demonstrated to improve solar cell performance and might possibly be reduced even more for closer packable non-spherical geometries formed under the influence of halogen compounds or halides during synthesis.[23,46]

## ASSOCIATED CONTENT

**Supporting Information.** Further experimental details, (HR)-TEM images, XRD diffractograms, UV/VIS and TXRF data as well as SEM images are provided. This material is available free of charge via the internet at http://pubs.acs.org.

## AUTHOR INFORMATION

### Corresponding Author

* meyns@chemie.uni-hamburg.de,
  beatriz.hernandez@uam.es,
  klinke@chemie.uni-hamburg.de

### Present Addresses

†Institute of Physical Chemistry, University of Hamburg, Grindelallee 117, 20146 Hamburg, Germany

## ACKNOWLEDGMENT

MM and CK thank the German Research Foundation DFG (Project number: KL 1453/5-1) and the European Research Council (Seventh Framework Program FP7, Project: ERC Starting Grant 2D-SYNETRA) for funding. Furthermore, the authors thank the Helmholtz Zentrum Berlin to have granted access the BESSY-II facility, in particular the SurICat UHV XPS station and Daniela Weinert for HR-TEM images. BHJ and RO thank for financial support from the Ministerio de Ciencia e Innovación (FIS2010-18847, FIS2012-33011 and Consolider-Ingenio en Nanociencia Molecular, ref CSD2007-00010), and the EU (SMALL, PITN-GA-2009-23884).